\documentclass[preprint2]{aastex}

\newcommand{\cmpss}{cm~s$^{-2}$}
\newcommand{\mps}{m~s$^{-1}$}
\newcommand{\kps}{km~s$^{-1}$}
\newcommand{\mpsini}{$M_p{\rm sin}i$}
\newcommand{\Msun}{${\rm M_\odot}$}
\newcommand{\Rsun}{${\rm R_\odot}$}
\newcommand{\Mjup}{${\rm M_J}$}
\newcommand{\xonb}{XO-3b}
\newcommand{\xon}{XO-3}
\newcommand{\Rjup}{${\rm R_J}$}
\newcommand{\vMs}{1.41}		
\newcommand{\eMs}{0.08}
\newcommand{\vRs}{2.13}		
\newcommand{\eRs}{0.21}
\newcommand{\sptype}{F5V}
\newcommand{\vrvK}{1471}		
\newcommand{\ervK}{48}
\newcommand{\vvsini}{18.54}
\newcommand{\evsini}{0.17}
\newcommand{\vfeh}{-0.177}
\newcommand{\efeh}{0.027}
\newcommand{\vDs}{260}		
\newcommand{\eDs}{23}
\newcommand{\vjd}{2454025.3967}	
\newcommand{\ejd}{0.0038}	
\newcommand{\vap}{0.0476}	
\newcommand{\eap}{0.0005}
\newcommand{\vperiod}{3.1915426}	
\newcommand{\eperiod}{0.00014}
\newcommand{\vecc}{0.260}
\newcommand{\eecc}{0.017}
\newcommand{\vMp}{13.02}		
\newcommand{\eMp}{0.64}	
\newcommand{\vMptot}{13.25}
\newcommand{\eMptot}{0.64}
\newcommand{\vRp}{1.95}		
\newcommand{\eRp}{0.16}
\newcommand{\vincl}{79.32}	
\newcommand{\eincl}{1.36}

\slugcomment{Submitted for publication in the Astrophysical Journal}

\begin{document}

\title{XO-3b: A Massive Planet in an Eccentric Orbit Transiting an \sptype\ Star}

\author{
Christopher~M.~Johns--Krull\altaffilmark{1,2},
Peter~R.~McCullough\altaffilmark{3},
Christopher~J.~Burke\altaffilmark{3},
Jeff~A.~Valenti\altaffilmark{3},
K.~A.~Janes\altaffilmark{4}, 
J.~N.~Heasley\altaffilmark{5},
L.~Prato\altaffilmark{6},
R.~Bissinger\altaffilmark{7},
M.~Fleenor\altaffilmark{8},
C.~N.~Foote\altaffilmark{9},
E.~Garcia--Melendo\altaffilmark{10},
B.~L.~Gary\altaffilmark{11},
P.~J.~Howell\altaffilmark{4},
F.~Mallia\altaffilmark{12},
G.~Masi\altaffilmark{13},
T.~Vanmunster\altaffilmark{14}
}

\email{cmj@rice.edu}

\altaffiltext{1}{Dept. of Physics and Astronomy, Rice University, 6100 Main Street, MS-108, Houston, TX 77005}
\altaffiltext{2}{Visiting Astronomer, McDonald Observatory, which is operated by the University of Texas at Austin.}
\altaffiltext{3}{Space Telescope Science Institute, 3700 San Martin Dr., Baltimore MD 21218}
\altaffiltext{4}{Boston University, Astronomy Dept., 725 Commonwealth Ave.,
Boston, MA 02215}
\altaffiltext{5}{University of Hawaii, Inst. for Astronomy, 2680 Woodlawn Dr., Honolulu, HI 96822}
\altaffiltext{6}{Lowell Observatory, 1400 West Mars Hill Road, Flagstaff, AZ,
86001}
\altaffiltext{7}{Racoon Run Observatory, 1142 Mataro Court, Pleasanton, CA 94566}
\altaffiltext{8}{Volunteer Observatory, 10305 Mantooth Lane, Knoxville, TN 37932}
\altaffiltext{9}{Vermillion Cliffs Observatory, 4175 E. Red Cliffs Drive, 
Kanab, UT 84741}
\altaffiltext{10}{Esteve Duran Observatory, El Montanya, Seva, 08553 Seva, Barcelona, Spain}
\altaffiltext{11}{Hereford Arizona Observatory, 5320 E. Calle Manzana, Hereford, AZ 85615}
\altaffiltext{12}{Campo Catino Astronomical Observatory, P.O. BOX 03016,
Guarcino (FR) Italy}
\altaffiltext{13}{Bellatrix Observatory, Via Madonna de Loco, 47 03023 
Ceccano (FR) Italy}
\altaffiltext{14}{CBA Belgium Observatory, Walhostraat 1A, B-3401 Landen, Belgium}

\begin{abstract}
We report the discovery of a massive (\mpsini $= \vMp \pm \eMp$ \Mjup;
total mass $\vMptot \pm \eMptot$\ \Mjup),
large ($\vRp \pm \eRp$\ \Rjup) planet in
a transiting, eccentric orbit ($e = \vecc \pm \eecc$) around a 10$^{\rm th}$
magnitude \sptype\ star in the constellation Camelopardalis.  We designate the 
planet \xonb, and the star \xon, also known as GSC 03727-01064.  
The orbital period of \xonb\ is $\vperiod \pm \eperiod$ days. 
\xon\ lacks a trigonometric distance; we estimate its 
distance to be \vDs$\pm$\eDs\ pc.
The radius of \xon\ is \vRs$\pm$\eRs\ \Rsun, its mass is \vMs$\pm$\eMs\ \Msun,
its $v$sin$i = \vvsini \pm \evsini$ \kps, and its metallicity is 
[Fe/H] $= \vfeh \pm
\efeh$.  This system is unusual for a number of reasons.  \xonb\ is one of 
the most massive planets discovered around any star for which the orbital 
period is less than 10 days. The mass is near the deuterium burning limit of
13 \Mjup, which is a proposed boundary between planets and brown dwarfs. 
Although Burrows et al. (2001) propose that formation in a disk or formation
in the interstellar medium in a manner similar to stars is a more logical way
to differentiate planets and brown dwarfs, our current observations are not 
adequate to address this distinction.  \xonb\ is also unusual in that its 
eccentricity is large given its relatively short orbital period.
Both the planetary radius and the inclination
are functions of the spectroscopically determined stellar radius.
Analysis of the transit light curve of \xonb\ suggests that the
spectroscopically derived parameters may be over estimated.  Though
relatively noisy, the light
curves favor a smaller radius in order to better match the steepness of
the ingress and egress.  The light curve fits imply a planetary
radius of $1.25 \pm 0.15$ \Rjup, which would correspond
to a mass of $12.03 \pm 0.46$ \Mjup.  A precise trigonometric parallax
measurement or a very accurate light curve is needed to resolve the 
uncertainty in the planetary mass and radius.
\end{abstract}


\section{Introduction}

There are now over 200 extrasolar planets known (http://exoplanets.eu/),
and most of these have been discovered using the radial velocity technique.
As a result, most of these systems yield the minimum mass of the planet
(\mpsini), its orbital semi-major axis, and properties of the central
star.  The data to date have dramatically changed our appreciation of
the diversity of planetary systems that can form, and have particularly
focussed the community's attention on the role that planetary migration
plays in the planet formation process (see review by Papaloizou et al. 2007).
As the precision and duration of radial velocity surveys increases, lower
mass planets and planets in longer orbits continue to be found.  As a
result, it has been suggested that the fraction of stars hosting
planets may be as high as 50\% (see Udry et al. 2007).  Therefore, it
appears that the planet formation process is relatively efficient.

The generally accepted model of giant planet formation is that of core
nucleated accretion (e.g., Pollack et al. 1996; Bodenheimer et al. 2000;
Hubickyj et al. 2005) in which gas giant planets first form a several
($\sim 10$) earth mass core of solids via non-elastic collisions in a disk
followed by the runaway accretion of gas from the disk once the mass of 
the core is sufficient to exert a strong gravitational influence.  Such a
model naturally predicts a positive correlation between the metallicity of 
the disk (and parent star) and the ease with which planets can form: there 
are more building blocks to form the solid core in higher metallicity cases.
The expected correlation has been suspected for some time (e.g., Gonzalez
1997; Santos et al. 2003), and has now been demonstrated in an unbiased way
(e.g., Santos et al. 2004; Fischer \& Valenti 2005) and has been taken as
evidence in support of the core accretion model of planet formation.  The
competing model, gravitational instability in a 
massive circumstellar disk (e.g., Boss 1997, 2000; Mayer et al. 2002) does 
not predict such a relationship.  Livio and Pringle (2003) showed
that lower disk metallicity reduces the efficiency of Type II migration
which may account for some of the increased
probability for finding a planet around more metal rich stars.  There is
some observational evidence suggesting a higher frequency for planets at
smaller separations around metal-rich stars 
which may support such a metallicity--migration
relationship (Jones 2004; Sozzetti 2004).  As a result, the 
planet--metallicity correlation may (at least in part) be due to a 
metallicity--migration relationship and may not simply signify a 
metallicity--formation relationship.

Additional observations of extrasolar planets are necessary to distinguish
between the competing modes of planet formation and the conditions which
affect planet migration scenarios.  For example, the core accretion model
requires several Myr to form Jovian planets in a disk (Inaba et al. 2003),
while gravitational instabilities could potentially form these planets around
very young stars.  Jovian planets are suspected around a few $\sim 1$ Myr
young stars (e.g., CoKu Tau/4 -- Forrest et al. 2004; GQ Lup -- Neuh\"auser
et al. 2005); however, clear detection and firm mass determinations from
techniques such as radial velocity variations remain elusive on such
young stars.  Transiting extrasolar planets offer the opportunity to determine
both the mass and radius of the planet, and hence the planet's density.
Such planets allow us to better constrain the variety of extrasolar planet
properties and offer another potential way to distinguish between planet 
formation scenarios.  For example, the high
core mass derived for the transiting Saturnian mass planet in orbit around 
HD 149026 (Sato et al. 2005) has been interpreted as strong support for 
the core accretion model.  
There are now $\sim 20$ known transiting extrasolar planets 
(http://exoplanets.eu/).  Several of these planets have unexpectedly large
radii (see Figure 3 of Bakos et al. 2007 and discussion therein), perhaps
suggesting additional heating of these planets other than that from 
irradiation by the star they orbit.  Additional effort is needed to
understand the structure of extrasolar planets and their origins.

Here, we report the discovery of the third transiting planet from the 
XO Project.
The XO Project aims to find planets transiting
stars sufficiently bright to enable interesting follow up
studies (McCullough et al. 2005).  \footnote{This paper includes data taken
on the Haleakala summit maintained by the University of Hawaii,
the Lowell Observatory, the Hobby-Eberly Telescope, 
the McDonald Observatory of The University of Texas
at Austin, and five backyard observatories.}
This planet presents some very interesting properties:  it
has a mass of \mpsini $=\vMp \pm \eMp$ \Mjup\ and is in a short period (3.2 d), 
eccentric orbit ($e = \vecc \pm \eecc$) around the apparently single \sptype\
star GSC 03727-01064.  \xonb\ has a radius of $\vRp \pm \eRp$ \Rjup, making
it substantially larger in both radius and mass than most of the other reported transiting
planets.
In \S 2 we discuss the observations leading to the discovery of \xonb.
In \S 3 we present our analysis of these observations to determine the
stellar and planetary properties.  In \S 4 we give a discussion of \xonb\
in the context of other extrasolar planets and current planet formation
models, and \S 5 summarizes our conclusions.

\section{Observations}

\subsection{XO Project Photometry}

\begin{figure}
\plotone{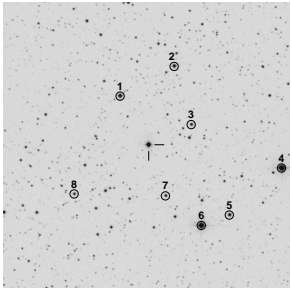}
\caption{\xon\ is centered, indicated by the two hash marks.
Stars from Table \ref{table:allsky} are circled. 
North is up; East to the left. The DSS image, digitized from a
POSSII-F plate with a IIIaF emulsion and an RG610 filter,
subtends 15\arcmin\ of declination.
\label{fig:allsky}}
\end{figure}

\begin{figure}
\plotone{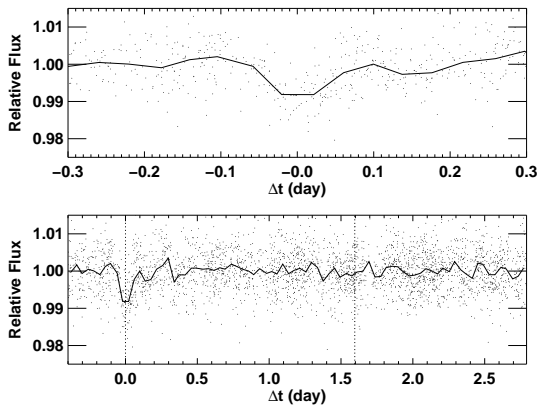}
\caption{A total of 2969 individual observations of \xon\ by the two XO cameras
over two seasons 2004 and 2005 are shown wrapped and phased according
to the transit ephemeris and averaged
in 0.01-day bins (line).  The top panel is the full lightcurve, and
the bottom panel shows the region around phase 0 (the primary transit,
marked with a vertical dashed line in the top panel)
enlarged.
From these data we identified the star
as a candidate for more refined photometry with other telescopes at
epochs of expected transits (Figure \ref{fig:etlc}).
\label{fig:xolc}}
\end{figure}

McCullough et al. (2005) describe the instrumentation, operation,
analysis, and preliminary results of the XO Project.  In summary, the XO
observatory monitored tens of thousands of bright ($V<12$) stars twice
every ten minutes on clear nights for more than 2 months per season
of visibility for each particular star, over the period September 2003
to September 2005. From our analysis of more than 3000 observations per
star, we identified \xon\ (Figure \ref{fig:allsky}) as one of dozens of stars
with light curves suggestive of a transiting planet. 
With the XO cameras on Haleakala, we observed four transits or 
segments of transits of \xon\ 
in 2003 and three in 2004, on Julian dates 2452934, 2452937, 2452998, 
2453001, 2453301, 2453320, and 2453352.
Table~\ref{table:lc} provides a sample of the photometry for \xon\
from the XO cameras.  The full table is available in the online
edition.  
From the survey photometry of \xon\ (Figure \ref{fig:xolc}),
which has a nominal standard deviation of 0.8\% or 8 mmag
per observation, we determined a preliminary light
curve and ephemeris, which we used to schedule
observations of higher quality with other telescopes,
as described in the next subsections. 

\subsection{Additional Photometry}

As outlined by McCullough and Burke (2006), once an interesting candidate
is detedcted in the XO photometry, the candidate is released to an Extended
Team (E.T.) of sophisticated amateur astronomers for additional observation.
In 2006 September through November and 2007 January through March, transit
events of \xon\
were observed by members of the E.T. from a total of 4 backyard 
observatories. 
Table~\ref{table:lc} also
provides E.T. photometry for \xon.  For the E.T. light
curves, the median differential magnitude out of transit provides the
flux normalization and the standard deviation out of transit provides
the uncertainty in the measurements.
A fifth E.T. observatory was used to obtain all-sky
photometry for \xon. These observatories are equipped with telescopes of 
aperture $\sim 0.3$ m.  The telescopes, equipped with CCD detectors,
are suitable for obtaining light curves sufficient to confirm the nature
of the transit and obtain good timing information.  A network of such
telescopes is well suited to observe candidates with known positions and 
ephemerides.  Here, we use two such telescopes in North America and two
in Europe.  Unlike the case of XO-1b (McCullough et al. 2006) and XO-2b
(Burke et al. 2007), we were not able to schedule observing time on a 
larger ($\sim 1$-m) telescope to obtain additional photometry -- the
E.T. photometry is the best we have obtained to date on \xonb.

We estimate all-sky photometric $B, V, R_{\rm C}$, and $I_{\rm C}$ magnitudes
for \xon\ and several nearby reference stars (Figure \ref{fig:allsky} and 
Table \ref{table:allsky}) calibrated using a total of 6 Landolt areas 
(Landolt 1992).  Using the Hereford Arizona Observatory (E.T. member
BLG) 0.36-meter telescope on photometric nights 2006 
October 27 and 2006 November 5, we measured the fluxes of 48 and 18 Landolt
stars, respectively at an air mass similar to that for \xon\ and established
the zero points of the instrumental magnitudes and transformation equations 
for the color corrections for each filter and the CCD. 
The derived magnitudes for \xon\ differed on the two dates by 0.03 magnitudes
or less in all 4 colors. 
The $B, V, R_{\rm C}$, and $I_{\rm C}$ absolute photometric accuracies
are 0.03, 0.03, 0.03, and 0.05 mag r.m.s.,
including both the formal error and an estimated systematic error. 
The Tycho magnitudes for \xon\ listed in 
Table \ref{table:star} transform (via Table 2 of Bessel 2000)
to Johnson $V = 9.86$, i.e. 0.06 mag (2-$\sigma$) fainter than our estimate.

\subsection{Spectroscopy}

In order to measure the orbital elements of the system, and in particular
the mass ratio, as well as to determine the characteristics of the host star,
we obtained spectra of \xon\ with two-dimensional cross-dispersed echelle 
spectrographs (Tull et al. 1995; Tull 1998) at the coude focus of the
2.7-m Harlan J. Smith (HJS) telescope and via a fiber optic cable on the 11-m
Hobby-Eberly Telescope (HET).  Both telescopes are located at McDonald 
Observatory.  The HJS spectra were obtained in a traditionally scheduled
manner, while the HET spectra were obtained in queue scheduled mode.
An iodine gas cell is used on the HET to provide the wavelength reference for
velocity determination.  While an iodine cell is available for HJS telescope,
it is not a facility instrument and the observations of \xon\ collected
here were done as a part of a radial velocity survey of young stars (Huerta
2007; Huerta et al. 2007) for which very high (few m s$^{-1}$) velocity
precision is not required.  Wavelength and resulting velocity calibration
is accomplished by taking thorium-argon reference lamp spectra before and
after each observation of \xon\ at the HJS telescope.  At the HJS we 
obtained one spectrum per night, and at the HET 
we obtained or two spectra per night.  At both telescopes, the 
spectral resolution was $R \equiv \lambda/\Delta\lambda 
\approx 60000$, and data were obtained on a total of
21 nights.  The two-dimensional echelle
spectra were reduced using IDL procedures described in Hinkle et al. (2000)
which include bias subtraction, flat fielding using a quartz lamp spectrum,
and optimal extraction of the data.  Table \ref{table:rvobs} gives a log of
the spectral observations.

\section{Analysis}

\subsection{Ephemeris}

We follow the same procedure used by Burke et al. (2007) for XO-2b to
refine the ephemeris of \xonb.  
We adopt the transit midpoint calculated for the event of 2006 October 16
as our ephemeris zeropoint: 2454025.3967$\pm$0.0021.  All measured
transit midpoints used to refine the ephemeris are reported in Table
\ref{table:ttime}.
To determine the orbital period, we minimize the $\chi^{2}$ difference between 
the observed transit times and a constant-period ephemeris model.
Based upon 3 transit events observed either in multiple passbands by the 
same observer, or observed by different observers,
we estimate the $1-\sigma$ uncertainty in the time of the center of an
individual transit to be 5 minutes, which we adopt as the
uncertainty for each of the transits observed in a single bandpass
by only one observer.
The best-fit period is \vperiod$\pm$\eperiod days.

\subsection{Radial Velocity Measurements}

Using spectra obtained at the HET, we measured \xon's radial velocities with
respect to the topocentric frame using iodine absorption lines superposed on
the spectra of \xon.  We modeled the extracted spectra using a template
stellar spectrum and the absorption spectrum of the HET iodine gas 
cell (Cochran 2000).  For the analysis of XO-1 (McCullough et al. 2006) and
XO-2 (Burke et al. 2007), a high resolution spectrum of the Sun and the 
Earth's atmosphere (Wallace et al. 1998) was used for the template stellar
spectrum.  We followed this same procedure initially for \xon; however,
the resulting radial velocity uncertainties were substantially higher than
we achieved for XO-1 and XO-2.  We suspect that the higher $v$sin$i$ of 
\xon\ compared to these other two stars contributes to this increased
uncertainty, but we were also concerned that the higher temperature of
\xon\ relative to these stars (and the Sun) resulted in spectral differences
large enough to increase the uncertainty further.  Therefore, we repeated
the HET radial velocity determinations using a high resolution 
($\lambda/\delta\lambda \sim 60000$) spectrum of the F5V star HD 30652 
from the SPOCS sample of stars (Valenti \& Fischer 2005, hereafter VF05) 
as the template
stellar spectrum.  The effective temperature of HD 30652 is 6424 K
(VF05), within 5 K of our derived $T_{eff}$ for \xon\
(see below), and $v$sin$i = 16.8$ km s$^{-1}$ which is close to the value
of 18.54 we find for \xon\ below.  The spectrum of HD 30652 is very similar
in appearance to that of \xon.  

Using an IDL\footnote{IDL is a software product of ITT Visual Information Solutions.}
implementation of Nelder and Mead's (1965) 
downhill simplex $\chi^2$ minimization algorithm, ``Amoeba,'' we adjusted
parameters of our model spectrum to fit the observations.
The model includes convolution of our model spectra with
a best-fitting Voigt profile to approximate the (slightly non-Gaussian)
line-spread function of the instrument.  The free parameters of our model
are a continuum normalization factor, the radial velocity of the star,
the radial velocity of the
iodine lines (which represent instrumental deviations
from their expected zero velocity with respect to the observatory),
and an exponent (optical depth scale factor) that scales the depths of
the lines as an arbitrary method of adjusting the spectrum of HD 30652
to even more closely match that of \xon.  Due to the iodine absorption, 
we could not
estimate the continuum level by interpolating between local maxima
in the spectrum, so instead we solved for the continuum iteratively,
as required to improve the fit between our model and the observations.
In the manner described above, for each $\sim$15 \AA\ section of each
individual spectrum within the region of the recorded spectrum with
significant iodine absorption, $5100 - 5700$ \AA,
we estimated the radial velocity of the star.  From the approximately
normal distribution of the resulting radial velocity estimates for each
epoch, we
calculated the mean radial velocity and its uncertainty.
The 1-$\sigma$ internal errors of the radial velocity measurements from
the HET spectra range from $\sim 120$ to $\sim 200$ \mps\ per epoch.  These
uncertainties are an order of magnitude larger than those achieved on 
XO-1 ($\sim 15$ \mps) and XO-2 ($\sim 20$ \mps).
We transformed our measured radial velocities to the barycentric frame
of the solar system and subtracted the mean radial velocity and report
these values in Table \ref{table:rvobs}.  These velocity measurements are
shown in Figure \ref{fig:rvfit}.

The HJS radial velocity measurement technique used here is described in 
detail in Huerta (2007) and Huerta et al. (2007).  We summarize it here.
The radial velocity shift of each observed spectrum is determined by a
cross correlation analysis of the observed spectrum with respect to a
reference spectrum.  To avoid complications which might result from a
spectral type mismatch, we use one particular HJS epoch (2454137.8215)
of \xon\ as the reference spectrum.  Again, wavelength calibration is
done by averaging the wavelength solution from Thorium-Argon reference
lamps taken before and after each stellar observation.  A total of 15
spectral orders which contain numerous, strong stellar lines and no
detectable telluric absorption lines are used in the cross correlation
analysis.  The radial velocity shift from the 15 orders are averaged
to get the final radial velocity shift for each observation, and the 
standard deviation of this mean is computed and adopted as the 
uncertainty of the cross correlation analysis.  Barycentric radial
velocity corrections are then applied to the observations to determine
accurate relative velocities for all the HJS spectra of \xon.

In addition to the uncertainty found above from the order-to-order
scatter in the HJS radial velocity measurements, there is an uncertainty
associated with the fact that our wavelength calibration is based on
spectra (Thorium-Argon) that are not observed simultaneously with the
stellar spectrum as is the case for iodine cell observations.  In order
to evaluate this uncertainty, a total of 6 stars from the Lick Planet
Search sample (Nidever et al. 2002; Butler et al. 1996; Cumming et al. 1999)
which are known to be stable at the 5 -- 20 \mps\
level are observed each night and analyzed in exactly the same way
as we treat the spectra of \xon.  The internal uncertainty in the radial
velocity shift measurements for these standard stars based on the measured
order-to-order scatter is typically $< 20$ \mps.  For the 2007 February
observing run (observations 3-10 at the HJS), the time series of radial 
velocity measurements for these standard stars show an average standard 
deviation of 123 \mps, with all but one of these stars showing a measured 
standard deviation less than 126 \mps.  Therefore, we adopt the average
value of 123 \mps\ as the uncertainty due to the non-simultaneous nature of 
the HJS wavelength calibration for the 2007 February HJS observations of
\xon\ and add it in quadrature to the uncertainty from the order-to-order
scatter for each observation.  For the 2006 October observations,
the radial velocity standard star data gives an uncertainty of 132 \mps\
due to the non-simultaneous nature of the target and wavelength calibration
spectra.  In addition, Huerta (2007) and Huerta et al. (2007) show that
there are systematic offsets of the order of 100 \mps\ in the radial
velocity standard star measurements from one observing run to the next, 
and speculate that this is caused by replacing the spectrometer slit plug
between observing runs (the slit plug is not moved during the observing
runs).  Using the radial velocity standard star data, we estimate that there
is a systematic shift of $140 \pm 33$ \mps\ between the 2006 October and 
2007 February observations.  We correct our 2006 October radial velocity
measurements of \xon\ by this amount and add this associated uncertainty
in quadrature with the other radial velocity uncertainties described
above.  The measured radial velocity shifts and the total uncertainty are 
reported in Table \ref{table:rvobs} and shown in Figure \ref{fig:rvfit} phased
to the ephemeris known from the transits.  The uncertainty for the radial 
velocity ($= 0$ \mps) of the reference epoch (2454137.8215) is set equal to 
the lowest uncertainty determined for the other HJS relative radial velocity
measurements.

An eccentricity approximately equal to zero is expected theoretically
for hot Jupiters in $\sim 3$ day orbits (Bodenheimer et al. 2001) and
was our expectation for \xonb.
Therefore, we expected sinusoidal radial velocity variations for \xon\ with
a phasing consistent with the transit observations.  The first two observations
obtained with the HJS, taken at phase $\sim 0.64$ and $\sim 0.95$ are
inconsistent with this expectation, forcing us to consider eccentric
orbits.  Many additional data points are required to adequately constrain an
eccentric orbit, so we performed intense observing of \xon\ with both the
HJS and the HET in late 2006 and early 2007.  When fitting the orbit, 
in addition to the radial velocity data, we also use the time of mid
transit as a constraint in the fitting.  Additionally, since the HJS and
HET data are on different relative velocity scales, we treat as a free
parameter the offset between these two scales.  In the orbit fitting,
we keep as fixed the orbital period and phases determined from the transit
ephemeris, and treat as free parameters the center of mass velocity of the
system, the eccentricity, the velocity amplitude $K$ for \xon, the longitude
of periastron, the phase of periastron passage, and the offset between 
the HJS and HET radial velocity measurements.  We use the nonlinear least 
squares technique of Marquardt (see Bevington \& Robinson 1992) to find the 
best fit parameters for the orbit, which are given in Table \ref{table:planet}.
Figure \ref{fig:rvfit} shows the radial velocity data along with the
orbital solution.  In this plot, the HET data are shifted (by adding
1192.52 \mps\ to the HET radial velocities) to match the 
HJS spectra and the center of mass velocity (1002.68 \mps) of all points is 
subtracted
for display purposes, since our center of mass velocity is based largely on
only relative data.  (As a result of these velocity shifts, the HJS velocities
in Table \ref{table:rvobs} have had 1002.68 \mps\ subtracted from them for 
display in Figure \ref{fig:rvfit}, and the HET velocities in Table 
\ref{table:rvobs} have had 189.84 \mps\ added to them for display in Figure
\ref{fig:rvfit}.)  Uncertainties in the orbital fit parameters are derived
by Mont\'e Carlo simulation of the data: for 1000 simulations we construct
fake radial velocity data using the orbital fit and applying Gaussian
random noise at a level commensurate with the observed data, and then fit this
model data using the same procedure outlined above.  Because the orbital
fit produces a minimum reduced $\chi^2 = 0.53$, we believe our radial velocity
uncertainty estimates may be a little too large.  This can be seen in
Table \ref{table:rvobs} where we report the observed minus calculated
($O-C$) velocities of the fit.  These values are typically less than our
derived uncertainties for the radial velocity measurements.  The standard
deviation of the HJS $O-C$ values is 122 \mps\ compared to a mean 
derived uncertainty of 156 \mps.  For the HET data, the standard deviation of
the $O-C$ values is 90 \mps compared to mean derived uncertainty of 169
\mps (though this is dominated by a single observation), indicating that the 
derived uncertainties for both data sets are too large.  Reducing these 
uncertainties would
reduce the uncertainties in the fit parameters, but we leave them as given,
since we can not justify any specific reduction in our measured radial
velocity uncertainties. From the fit, we find the velocity 
semi-amplitude of the orbit of \xon\ is K = \vrvK$\pm$\ervK\ \mps.

As described below, the $v$sin$i$ of \xon\ is relatively large ($18.5 \pm 0.2$
\kps), which is substantially larger than most stars around which planets have
been discovered (HAT-P-2b is another example of a planet orbiting a star which
has a relatively large $v$sin$i = 19.8 \pm 1.6$ \kps, Bakos et al. 2007).  
The experience of Mandushev et al. (2005) and the similarity between the
primary in that system with \xon\ gives concern that \xon\ could be a 
hierachical triple system in which two lower mass stars comprise a short
period eclipsing binary star which are themselves orbiting a more massive, 
significantly brighter primary in a much longer period orbit.  In the case 
of Mandushev et al. (2005), the short period binary stars
display tens of \kps\ velocity shifts, but the strong, broad lines of the
system almost hide this signal.  The apparent velocity shift of the system
primary is then the result of weaker lines moving around the (stationary) lines
of the primary, producing line profile distortions which are misinterpreted 
as radial velocity variability of the primary.

Line bisector analysis is a standard technique to look for evidence that line
distortions are producing false radial velocity variations (e.g., Hatzes et
al. 1997; Martinez Fiorenzano et al. 2005), and have been used successfully 
to identify false radial velocity variations caused by star spots (e.g.,
Queloz et al. 2001; Bouvier et al. 2007; Huerta et al. 2007) and 
hierarchical triple systems 
(e.g., Santos et al.  2002; Mandushev et al. 2005).  To improve signal-to-noise
in the bisector analysis, it is common practice to compute the bisector of 
the spectrum cross correlation function (Queloz et al. 2001; Santos et al. 
2002; Mandushev et al. 2005).  Here, we calculate the line bisector of the 
total cross correlation function computed from all 15 orders in the HJS 
spectra used in the determination of the radial velocities.  We then compute
the bisector span, which is the difference of the line bisector at two
reference levels.  Here we define the bisector span, $S_B$, as the value
of the bisector at an absolute cross correlation level of 0.15 minus the 
bisector at an absolute cross correlation level of 0.75.  These span values
are given in Table 
\ref{table:rvobs}.  Because the spectral regions of the HET data 
which contain strong stellar lines also contain numerous iodine absorption 
lines, we restrict ourselves to the HJS data for the analysis of line
bisectors.  Previous studies which used bisector analysis to reveal a false
positive in planet searches find a clear correlation between the measured
radial velocity and the bisector span (Queloz et al. 2001; Santos et al.
2002; Mandushev et al. 2005; Bouvier et al. 2007; Huerta et al. 2007).
In these cases, 
whether due to spots (e.g., Queloz et al. 2001; Bouvier et al. 2007; Huerta
et al. 2007) or due to a hirearchical triple system (Santos et al. 2002; 
Mandushev et al. 2005), the full range measured in $S_B$ is approximately 
the same as the full range of the measured radial velocities.  

On the other hand, Torres et al. (2004) use the lack of a correlation between
$S_B$ and the radial velocity to help confirm a planet transiting OGLE-TR-56b.
In addition, for OGLE-TR-56b the full range in $S_B$ is less than one third 
the full range in the radial velocities, which Torres et al. (2004) argue is 
additional evidence for the planetary nature of this object.  
Figure \ref{fig:bisect} shows the bisector span,
$S_B$, plotted versus the measured radial velocity for the HJS spectra of
\xon.  Also given in the figure is the value of the linear correlation
coefficient and its associated false alarm probability (Bevington \&
Robinson 1992) for these data.  In addition to there being no 
significant correlation
between $S_B$ and the radial velocity, the full range of $S_B$ is very
small compared to the full range in the radial velocity.  
Figure \ref{fig:bisect} and Table \ref{table:rvobs} show that
the full range in $S_B$ in \xon\ is less than one
tenth the full range in the radial velocity variations for \xon.

As a last check that the photometric dimmings and radial velocity variations
of \xon\ are due to a planetary mass companion and not the result of a
hierarchical triple system with later type stars present, we looked for an
infrared (IR) excess in \xon.  O'Donovan et al. (2006) used a redder than 
expected $V-K$ measurement as one piece of evidence to reject the planetary
hypothesis for the F star GSC 03885-00829.  Using the photometry reported in
Table \ref{table:star}, we estimate the color $V-K_{\rm s} = 1.01 \pm 0.06$ 
for \xon.  Based on the effective temperature and gravity found below for 
\xon, we assign a spectral type of \sptype\ based on the calibrations presented
in Cox (2000).  In the color system of Bessell and Brett (1988),
an F5V star such as XO-3 is expected to have an intrinsic 
$(V - K_{\rm s})_\circ = 1.10$. For comparison to the 2MASS colors reported in
Table \ref{table:star}, we use the color transformation relations found
in the {\it Explanatory Supplement to the 2MASS All Sky Data Release and 
Extended Mission Products} by Cutri et al. 
(http://www.ipac.caltech.edu/2mass/releases/allsky/doc/), and find that we
must add 0.039 mag to the Bessel \& Brett value in 
order to estimate $(V - K_{\rm s})_\circ$ = 1.14 on the 2MASS system. Thus, 
the observed $V-K_{\rm s} = 1.01$ color of XO-3 is 0.13 mag bluer 
($\sim2$-$\sigma$) than its expected color.
The slight bluing may be the result of the
relatively low metallicity we derive (see below), but there certainly is
no evidence for a near IR excess suggesting a hierarchical triple system.
The bisector analysis and the $V-K_{\rm s}$ color of \xon\ give us confidence
that both the photometric dimming and radial velocity variations we measure 
in this star are indeed due to a planetary mass companion.

\subsection{Spectroscopically-Derived Stellar Properties and Planetary Mass}

We used the software package SME (Valenti \& Piskunov 1996) to fit each
of the 10 spectra of \xon\ from the HJS telescope with synthetic spectra.
We used
the methodology of VF05, including their minor
corrections to match the Sun and remove abundance trends with temperature
(negligible in this case). Because of gaps between echelle orders,
the HJS spectra are missing the wavelength intervals 6000--6123
\AA, which was included in the Valenti \& Fischer
analysis. These wavelength intervals are also missing from our extracted
HET spectra because the relevant echelle orders span the gap between
the two detectors.

We averaged our SME results for the 10 HJS spectra, obtaining
the parameter values in Table \ref{table:sme}. 
Each value in the last column of the table, labeled ``Precision'' because
systematic uncertainties are not included,
is the standard deviation of the 10 measurements divided
by $\sqrt{9}$ to yield the formal uncertainty in the mean.
The median value of
each derived parameter (not given) differs from the mean by less than the
uncertainty in the mean. The final row in the table gives [Si/Fe], which
VF05 used as a proxy for alpha-element enrichment,
when interpolating isochrones.
Figure \ref{fig:mcd} shows \xon's spectrum in the region of the \ion{Mg}{1}
B triplet, which is the dominant spectroscopic constraint on gravity. These
three Mg lines also have a significant impact on the global [M/H]
parameter, which is used to scale solar abundances for all elements
other than Na, Si, Ti, Fe, and Ni.

Following the methodology of Fischer \& Valenti (2005), we interpolated
Yonsei-Yale (Y$^2$) isochrones (Demarque et al. 2004) to determine
probability distribution functions for the mass, radius, gravity, and age
of \xon. The trigonometric parallax of \xon\ is unknown, so we initially
assumed distances of 240, 260, and 280 pc with an adopted uncertainty of
10 pc in each case (which affects the width of the resulting distribution
functions).  In order to perform the isochrone analysis, we need a $V$ magnitude
corrected for extinction for \xon.  We measured $V = 9.80\pm0.03$
and $B = 10.25\pm0.03$ for \xon\ (Table \ref{table:star}).  These
observations give a measured $B-V = 0.45\pm0.04$ for \xon.
VandenBerg \& Clem (2003) determined empirical color-temperature relations
for stars including variations with gravity.  We downloaded their high
temperature table (mentioned in footnote 1 of their paper) and 
interpolated in the table to the effective temperature, surface gravity, 
and metallicity of \xon\ determined above.  Doing so gives a predicted
$(B-V)_\circ = 0.440$ for \xon.  As discussed below, the lightcurve
analysis and the derived properties of \xonb\ favor a larger stellar
gravity for \xon.  So, we also interpolate the predicted colors to a
gravity of log$g = 4.2$ (a little more than 3$\sigma$ higher than 
the spectroscopically derived value).  Doing so results in a predicted 
$(B-V)_\circ = 0.446$.  These two predicted values are within
0.01 and 0.001 mag of the measured $B-V$ of \xon, with an
uncertainty of 0.04 in the measured color.  We therefore find no
compelling evidence for significant reddening to \xon\ and we adopt
$V = 9.80\pm0.03$ as the extinction corrected magnitude of the star.

We used our spectroscopic effective temperature, spectroscopic gravity,
and an assumed distance to derive a bolometric correction by interpolating
the ``high temperature'' table from VandenBerg \& Clem (2003).
We combined the bolometric
correction with the observed V-band magnitude to determine stellar
luminosity. Then we used the stellar luminosity and our spectroscopic
effective temperature, iron abundance, and alpha-element enrichment to
interpolate the Y$^2$ isochrones to produce the probability distribution 
functions in Figure \ref{fig:stellarparam}. 

The parameters we derive place \xon\ above the zero-age main sequence.  As 
a result, reasonable fits to the data are possible for pre-main sequence 
evolutionary tracks.  However, if \xon\ were indeed a pre-main sequence star,
we would expect strong \ion{Li}{1} absorption at 6708 \AA.  For example, in 
their survey of Pleiades F stars, Boesgaard et al. (1988) show that the
\ion{Li}{1} line at 6708 \AA\ is approximately as strong as the nearby 
\ion{Ca}{1} line at 6718 \AA\ is stars of similar spectral type to \xon.  We 
show in Figure \ref{fig:licheck} the Li line region in \xon\ produced by 
averaging the spectra obtained at the HET (this region is free of iodine 
absorption).  The position of the \ion{Li}{1} line is marked and the nearby 
\ion{Ca}{1} line is clearly visible.  The Li line is not detectable, 
indicating that \xon\ is substantially older than the Pleiades, therefore we
conclude \xon\ is not a pre-main sequence star.

The most probable distance, mass, radius, and age for \xon\ are 260 pc, 
1.41 M$_\odot$, 2.13 R$_\odot$, and 2.69 Gyr respectively, for our best fit
gravity of log$g = 3.95$.  Combined with our measured semi-amplitude
K = \vrvK$\pm$\ervK\ \mps, this stellar mass implies a mass of 
$M_p$sin$i = $\vMp$\pm$\eMp\ \Mjup\ for the planet \xonb.
If we assume that \xon\ is at 240 pc instead, then the most probable 
mass, radius, age, and gravity are 1.36 M$_\odot$, 1.95 R$_\odot$, 2.78 Gyr,
and log$g = 4.00$, still placing the star well above the main sequence.
As discussed below, one of the unusual properties of \xonb\ is the relatively
large radius we derive for the planet.  The planetary radius is approximately
proportional to the inferred stellar radius.  The planetary radius could be
smaller if we could justify a smaller stellar radius; however, doing so begins 
to bring the inferred gravity in conflict with the spectroscopically derived
value of log$g = 3.95\pm0.062$.  Taking a 3$\sigma$ upper limit, the largest
allowed value for the gravity is log$g = 4.14$, corresponding to a stellar
mass of 1.27 M$_\odot$ and a stellar radius of 1.62 R$_\odot$ for an inferred
distance of 200 pc.  Together with the measured velocity semi-amplitude, 
these values imply a planetary mass of $M_p$sin$i = 12.14\pm0.40$ \Mjup\
where the uncertainty here is just that resulting from the radial velocity fit.
Clearly, a trigonometric parallax measurement for 
\xon\ will be of great value in better establishing both the stellar and 
planetary parameters.

\subsection{Light Curve Modeling and the Planetary Radius}

\begin{figure}
\plotone{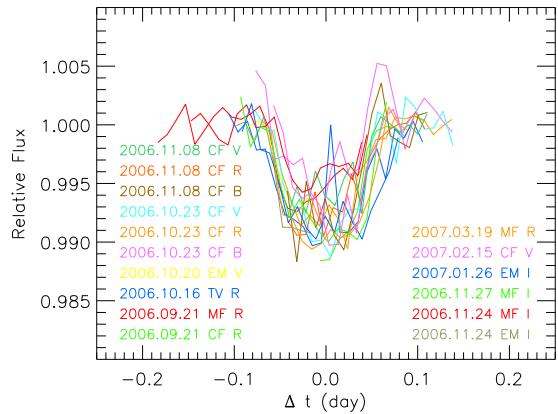}
\caption{Time series photometry of \xon\ from 2005 -- 2007, with dates, 
observers, and
filters indicated. The observations have been averaged in 0.006-day bins.
The figure is in color in the electronic edition.
\label{fig:etlc}}
\end{figure}

\begin{figure}
\epsscale{0.75}
\plotone{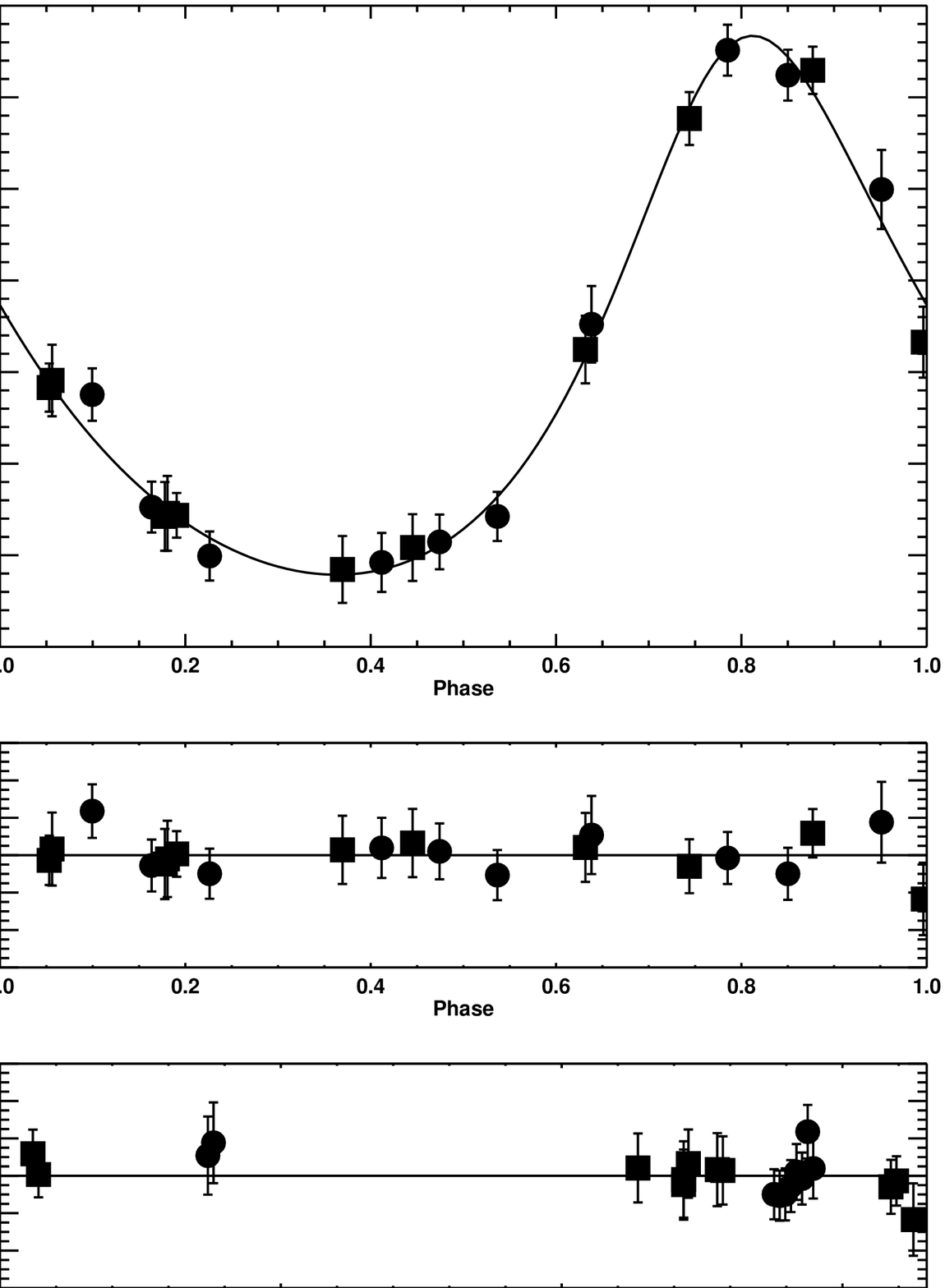}\\[5mm]
\caption{Top: The measured radial velocity data are shown along with the best fit
orbit.  The filled circles are based on data from the 2.7 m Harlan J. Smith
Telescope, and the filled squares come from data taken with the 11-m HET.
The radial velocity curve of \xon\ traces out an eccentric orbit
with a velocity amplitude K = \vrvK$\pm$\ervK\ \mps, implying \xonb's mass is
\mpsini $= \vMp\pm\eMp$\ \Mjup. The period and phase used to fold 
the measured radial 
velocities are fixed at values determined by the photometric transits.  The 
measured center of mass velocity with respect to the solar system's barycenter
has been subtracted (its value is arbitrary since it is largely based on the
intrinsically relative radial velocity measurements from the HJS).
Middle: A plot of the radial velocity residuals, again phase folded with
the ephemeris determined from the photometric transits.
Bottom:  The radial velocity residuals are shown again, this time as a 
function of time, showing no long term trend.
\label{fig:rvfit}}
\end{figure}

\begin{figure}
\plotone{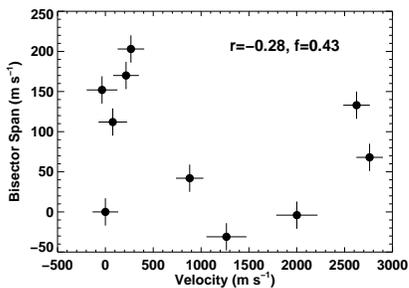}
\caption{The bisector span is plotted as a function of the measured
radial velocity for all the HJS spectra of \xon.  There is no correlation
of the span with radial velocity and the full amplitude of the span
variations is less than a tenth of the full amplitude of the radial
velocity variations.
\label{fig:bisect}}
\end{figure}

\begin{figure}
\epsscale{0.8}
\plotone{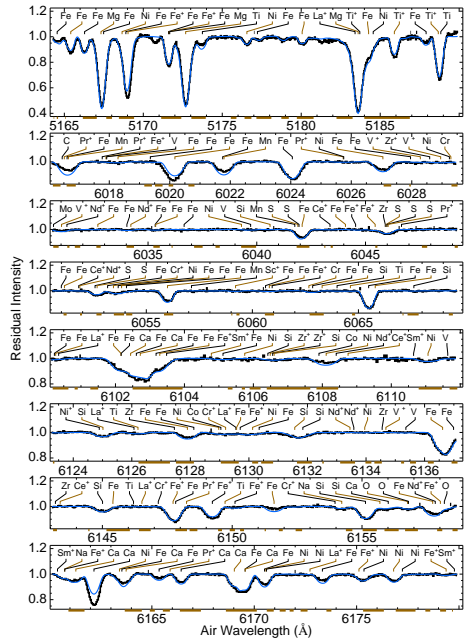}
\caption{
The mean spectrum of \xon\ as observed (black
histogram) and modeled with SME (curve, colored blue in the electronic edition)
in the region of the Mg b triplet.
Labels note the elements responsible for the indicated spectral lines.
Intermittent line segments (tan) beneath the horizontal
axis indicate wavelength intervals used to constrain the
spectroscopic parameters. Very short and intermittent line segments (black)
immediately above
the spectrum indicate wavelength intervals used to constrain the
continuum fit.  A color version of this figure appears in the electronic
version of the paper.
\label{fig:mcd}}
\end{figure}

\begin{figure}
\plotone{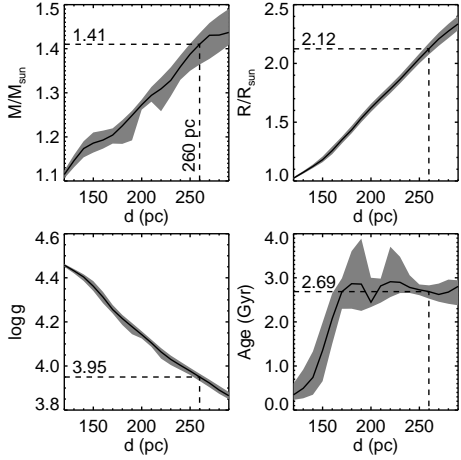}
\caption{The stellar mass, radius, gravity, and age which result from the
isochrone analysis.  The solid line shows the derived values as a function
of the assumed distance, and the gray region shows the 68\% confidence
limit on these parameters.  Dashed lines are drawn at 260 pc corresponding
to the spectroscopically determined log$g = 3.95$.
\label{fig:stellarparam}}
\end{figure}

\begin{figure}
\plotone{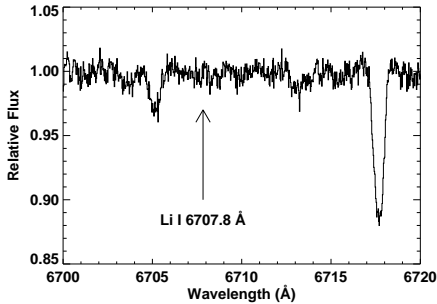}
\caption{The spectrum of \xon\ in the wavelength region of the \ion{Li}{1}
doublet at 6708 \AA.  The position of the Li feature is marked.  The Li
feature is expected to be approximately as strong as the \ion{Ca}{1} line
at 6718 \AA\ in Pleiades age stars of similar spectral type, and even
stronger at younger ages.  Therefore, we conclude \xon\ is not a 
pre-main sequence star, but instead is likely on the post-main sequence.
\label{fig:licheck}}
\end{figure}

\begin{figure}
\plotone{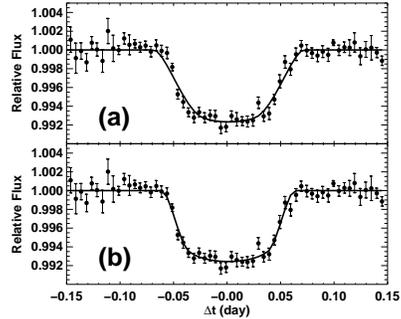}
\caption{The combined E.T. transit light curve of \xon\ and the best fit
transit light curve models computed under different assumptions.  The upper
panel (a) fixes the stellar mass and radius to the values determined from
the combined spectroscopic and isochrone analysis.  The lower panel (b) is
determined by minimizing $\chi^2$ in the light curve fit by letting the
stellar mass and radius vary according to the isochrone results for different
assumed distances.  This best fit produces a stellar gravity (log$g$) more 
than $3\sigma$ greater than the measured value from the spectroscopic
analysis.
\label{fig:lcurvefit}}
\end{figure}

For additional photometric
analysis and lightcurve fitting, all the E.T. photometric data shown in
Figure \ref{fig:etlc} were averaged into 7.4 minute bins.
Binning permitted
outlier rejection and empirical estimation of the noise by the scatter
of the individual observations, which is helpful in cases such as this
in which residual calibration errors can be significant.
Differences in the depth and the shape
of the transit light curve in different photometric bands are both expected and
observed to be
substantially smaller than the photometric uncertainties in the E.T.
lightcurves, so we combined the light curve from all photometric bands
to improve the signal to noise ratio.  The
binning begins with phasing the light curves
using the refined ephemeris determined in \S3.1.
The final binned light curve consists of a robust
average of E.~T.\ photometric measurements in bins 7.4 minutes in duration.
This robust average is determined by first calculating
the median and median absolute deviation for each bin and rejecting points
in the bin that deviate by more than $4\sigma$ from the median.  Next,
the mean and standard deviation are calculated for each bin and points
which are more than $4\sigma$ from the mean are rejected.  This rejection
based on the mean and standard deviation is repeated one more time before
a final (``robust") mean is calculated.
The final binned light curve is supplied in Table \ref{table:lc} and
is shown in Figure \ref{fig:lcurvefit} along
with the fits described below.  
The uncertainty in each binned light curve data point is the standard 
deviation of (surviving) measurements in the bin divided by the square root 
of the number of measurements in that bin.  The uncertainties in all bins are
then multiplied by a correction factor as described below (the uncertainties
in Table \ref{table:lc} do not have this correction factor applied).

We modeled the mean transit light curve using the analytic transit model
of Mandel and Agol (2002).  For a planet with an
eccentric orbit, the transit model has ten parameters: transit
midpoint ($t_{o}$), orbital period, stellar mass ($M_{\star}$),
stellar radius ($R_{\star}$), planet radius ($R_{\rm p}$), inclination
($i$), quadratic limb darkening law coefficients ($u_{1}$ and
$u_{2}$), eccentricity ($e$), and longitude of perihelion ($\omega$).
Throughout the following analysis, the orbital period remains fixed at
the period found in \S3.1.  The
radial velocity data provide the best estimates for $e=0.26$ and
$\omega=-15.4$ that remain fixed during the $\chi^{2}$ minimization.
For each fit presented below, we fix the stellar parameters to specific
values based on the spectral synthesis and isochrone analysis presented
above.
We interpolate quadratic limb darkening coefficients from Claret (2000) for 
the Cousins R photometric bandpass ($u_1 = 0.220$; 
$u_2 = 0.380$) based on the gravity, effective temperature, and 
metallicity ([M/H]) determined from the spectroscopic analysis
(Table \ref{table:sme}).
The remaining parameters,
$t_{o}$, $R_{\rm p}$, and $i$, are allowed to vary and are solved for
using the ``Amoeba" algorithm mentioned earlier.
For input into the Mandel and Agol (2002) transit model, we determine the
projected separation between star and planet, $\delta$, in the case of
nonzero eccentricity following the procedure outlined by
Hilditch (2001).  For a given $i$, $\omega$, and $e$, we determine the
true anomaly, $\theta{\rm min}$, that minimizes $\delta$ (Equation~4.9
of Hilditch 2001).  The resulting $\theta{\rm min}$ corresponds to the
mean anomaly at transit midpoint providing the zeropoint to convert
observed times to $\theta$ via Kepler's equation.

We then initially fit the light curve assuming the
stellar mass (1.41 \Msun) and radius (2.13 \Rsun) derived from the
spectral fits as described above.  The free parameters of the light curve
fit are then the radius of the planet and the inclination of the orbit
(and the time of mid-transit).  The resulting fit is shown
in the top panel of Figure \ref{fig:lcurvefit}.  We then repeated the light
curve fits using the extremes in the stellar mass (1.33 and 1.90 \Msun)
and radius (1.49 and 2.36 \Rsun) which result from the $1\sigma$ uncertainties
in these parameters derived above.  Together, this then gives us estimates
of the planetary radius ($R_p = \vRp \pm \eRp$ \Rjup) and orbital inclination
($i = 79.^\circ32 \pm 1.^\circ36$) which are also given in Table 
\ref{table:planet}.  The total mass of the planet is then
$\vMptot \pm \eMptot$ \Mjup.  Examination of the figure shows that the
model does not fit all aspects of the lightcurve: the model ingress and
egress appear noticeably longer than the observed ingress and egress. 

In order to better model the shape of the ingress and egress while still
maintaining the total duration of the transit, the radius of the star and
the planet both need to be reduced and the inclination must increase so that
the total chord length the planet traverses accross the face of the star
stays approximately the same.  Figure \ref{fig:stellarparam} shows that
from the isochrone analysis, the stellar mass and radius both decrease
as the distance is assumed to be smaller.  We have
performed the isochrone analysis every 10 pc, so we computed model
light curves, stepping down in distance 10 pc at a time from our favored
distance of 260 pc.  For each new distance, we adopt the corresponding stellar
mass and radius resulting from the isochrone analysis and compute $\chi^2$
for the fit to the light curve.  The minimum in $\chi^2$ occurs between 190
and 200 pc.  Fitting a parabola to the lowest 4 $\chi^2$ values, gives a
best fit to the light curve for a distance of 185 pc which gives 
$M_* = 1.24$ \Msun, $R_* = 1.48$ \Rsun, and log$g = 4.19$.  This fit is
shown in the lower panel of Figure \ref{fig:lcurvefit}.  As mentioned
earlier, the the best $\chi^2$ is quite large ($\sim 129$) for the
63 degrees of freedom in our fit, indicating that the photometric 
uncertainties have been underestimated.  We attempt to correct this by
determing the multiplicative factor required to give a $\chi^2 = 63.0$
(a reduced $\chi^2 = 1.0$).  The uncertainties shown in Figure 
\ref{fig:lcurvefit} have been multiplied by this factor.  Having done
this, we can then assign as the uncertainty in $R_P$ and $i$
the difference between the best fit parameters and those we obtain
when $\Delta\chi^2 = 1.0$.  Doing so,
we find that the transit lightcurve is best fit for $R_p = 1.25 \pm 0.15$
\Rjup\ and $i = 83.^\circ32 \pm 1.^\circ26$.  The total mass of the planet
is then $12.03 \pm 0.46$ \Mjup.  This value for the planetary
radius is considerably lower than the value derived above from the default
stellar parameters.  As discussed below, we regard this as a lower limit
to the true planetary radius; however, we note that the ambiguity 
described here can be greatly diminished by obtaining very accurate
photometry of a transit of \xon\ and/or by obtaining a precise trigonometric
parallax measurement.

\section{Discussion}

\subsection{Comparison to other Transiting Planets\label{sec:cf}}

The $\sim 13$ \Mjup\ planet \xonb\ is unusual in many respects compared to
the sample of known extra-solar planets.  The mass of \xonb\ is quite large 
compared to most of the known extra-solar planets.  Zucker and Mazeh (2002) 
noted that the most massive short period planets are all found in multiple
star systems.  Udry et al. (2003) emphasized the general lack of massive 
planets on short period orbits, particularly for planets orbiting single 
stars, and interpreted these results in terms of planetary migration 
scenarios.  The star \xon\ is not known to be a binary, though
it is relatively unstudied.  While it is in the Tycho-2 catalog
(H{\o}g et al. 2000), \xon\ does not have a significant proper motion measurement
from the {\it Hipparcos} mission.  It is not known if any of the nearby
stars seen in Figure \ref{fig:allsky} are common proper motion companions
or not.  Udry et al. (2003) point out that among planets orbiting single
stars known at that time, there were no planets more massive than 2 \Mjup\
with periods less than 100 d (see also Eggenberger et al. 2004).  Using 
data for 218 extra-solar planets as compiled on the Geneva Extrasolar Planet 
website (http://www.exoplanets.eu) updated as of 27 May 2007, this general 
lack of short period, very massive planets is still apparent.  There are only 3
other planets (HD 162020b, HD 17156b, and HAT-P-2b)
orbiting apparently single stars with periods less than 30 days
and masses larger than 2.5 \Mjup, and one of those (HD 162020b) is suspected
to actually be a much more massive brown dwarf (Udry et al. 2002).  There is
only 1 planet other than \xonb\ with a mass larger than 2.5 \Mjup\ and a
period less than 4 days (HD 120136b, M$_2$sin$i = 4.14$ \Mjup, Butler et al.
1997), and this star is known to be in a stellar binary system (Hale 1994).

The eccentricity of \xonb\ is also quite rare relative to other known
extra-solar planets, given the short orbital period of \xonb.  The 
eccentricity of the shortest period extra-solar planets are all quite low
($e < 0.05$) and most are consistent with zero eccentricity (Halbwachs et
al. 2005).  This is generally believed to be due to either tidal 
circularization (e.g., Wu 2003; Ivanov \& Papaloizou 2007) or the result of 
smooth orbital migration in a disk which is not thought to produce significant
eccentricity (Murray et al. 1998).  The eccentricity of \xonb\ is larger than
all other planets with periods less than that of \xonb.  
We discuss the
eccentricity of \xonb\ further in Section \ref{sec:tidal}.  While the mass, period, and 
eccentricity of \xonb\ make it quite rare among extra-solar planets, it
is not totally unique.  The recently discovered HAT-P-2b (Bakos et al. 2007)
is very similar in many respects with a period of 5.6 d, $M_p = 9.04$ \Mjup,
and $e = 0.52$.  
Also, the hot Neptune GJ 436b has a period of 2.6 days and e = 0.15
(Butler et al 2004; Deming et al 2007). Apparently the assumption,
commonly-held prior to 2007, that short-period planets should have
circular orbits was incorrect.

It is now well established that host star metallicity correlates with the
likelihood of finding a Jupiter mass companion orbiting the star in the
sense that higher metallicity stars are more likely to have planets (Santos
et al. 2004; Fischer \& Valenti 2005).  The metallicity of \xon\ ([M/H] 
= -0.20) is relatively low, and the above mentioned studies show that planets
are found around only $\sim 3$\% of stars with similar metallicity.
There are only two other very 
massive planets ($M_p > 5$ \Mjup) known around stars (HD 111232, HD 114762)
with metallicities lower than that of \xon.  The correlation between host
star metallicity and the probability of finding a planet is often taken as
support for the core accretion model (e.g., Pollack et al. 1996; Bodenheimer 
et al. 2000; Hubickyj et al. 2005) of planet formation.  Under that assumption
though, it may be surprising that a few very massive planets have also been
found around low metallicity stars.   It has also been suggested that 
host star metallicity decreases on average as the planet mass increases 
which appears inconsistent with the core accretion model (Ribas \&
Miralda--Escud\'e 2007).  On the other hand, gravitational instabilities in 
a disk are not expected to produce a metallicity-planet correlation 
(Boss 2002).  Taking this notion further, Ribas and Miralda--Escud\'e
(2007) have suggested that the sample of extra-solar planets consists of
objects formed via two different paths: core accretion in a disk and 
fragmentation of a pre-stellar cloud.  The high mass, relatively large
eccentricity, and low metallicity of the \xonb\ system would then ``fit''
the notion of Ribas and Miralda--Escud\'e of a planet
formed from the collapse of a pre-stellar cloud, but of course the
observations of \xonb\ do not prove (or disprove) either formation scenario.

\subsection{How Big is \xonb?}

Potentially, one of the most unusual and interesting properties of \xonb\ is
the large radius of the planet.  At an age of 2-3 Gyr, giant planets and
low mass brown dwarfs are expected to have radii very close to that of Jupiter
(Burrows et al. 2001; Baraffe et al. 2003).  Using the tables in Baraffe et 
al. (2003), the radius of a 13 \Mjup\ planet is expected to be 1.03 \Rjup\ at
1 Gyr, and 0.97 \Rjup\ at 5 Gyr.  Including heating from the central star can
increase the expected radius for so-called ``hot Jupiters" (e.g., Bodenheimer 
et al. 2003).  Fortney et al. (2007) compute planetary structure models up to 
11.3 \Mjup\ for a range of orbital separations, including the insolation
produced by absorption of solar radiation.  At an age of 300 Myr or older,
all the models computed have radii less than 1.3 \Rjup.  In all cases
analyzed by Fortney et al., their models
predict a substantially smaller radius than the $\vRp \pm \eRp$ \Rjup\ 
determined based upon the values of the stellar mass and radius inferred 
from the spectroscopic and isochrone analysis of \xon.  Such a large
discrepancy between the observed and predicted planetary radius is 
reminiscent of TReS-4 which has a radius of $1.674 \pm 0.094$ \Rjup\
(Mandushev et al. 2007).  However, the radius of \xonb\
is quite uncertain, and the light curve itself favors a smaller planetary 
radius of $R_p = 1.25 \pm 0.15$ \Rjup.  Because \xonb\ is in an eccentric 
orbit, for comparison with the models of Fortney et al. (2007)
we compute the average orbital
separation between \xonb\ and its star and then adjust this separation to
take into account the difference in luminosity between
\xon\ and the Sun.  The smaller planetary radius for \xonb\ is associated
with a smaller
stellar radius for \xon, so for this exercise, we take $R_* = 1.57$ \Rsun.
We thus compute an effective distance of 0.024 AU to use for \xonb\ when
using the tables of Fortney et al. (2007).  Interpolating in these tables,
the predicted radius for \xonb\ is 1.18 \Rjup\ at 1 Gyr and 1.11 \Rjup\ at
4.5 Gyr.  Both values are lower than the smaller radius inferred for \xonb,
but only by about $1\sigma$.  A more precise radius for \xonb\ would be
a very interesting 
benchmark for the theory of extra-solar planets.

The actual radius of \xonb\ can be better constrained by a true parallax
measurement and more precise photometric observations (ideally in multiple
colors) of additional transits.  The constraint from this work that favors the larger
radius for \xonb\ is the stellar gravity inferred from the spectroscopic
analysis.  A number of investigators have pointed out that stellar gravity
determinations are notoriously difficult and the relative error in this
parameter is often substantially larger than almost all other measured
properties of transiting extra-solar planetary systems.  Many of these
investigators have elected to use the transit light curve fits in combination
with stellar isochrone models to estimate the stellar and planetary radius
and hence log$g$ for the star (e.g., Sozzetti et al. 2004, 2007; O'Donovan 
et al.  2006; Bakos et al. 2007; Mandushev et al. 2007).  The photometry
of \xonb\ is not as precise as that used in those studies, so
we do not emphasize 
the photometrically-derived parameters as those investigators have. However, we can
examine the spectroscopically determined log$g$ to estimate by how much,
and in what sense, it is in error.  

\begin{figure}
\plotone{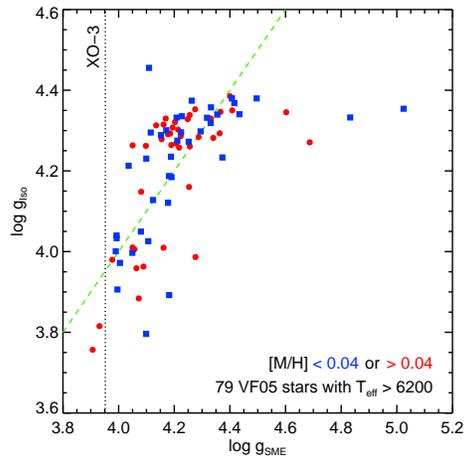}
\caption{Comparison of spectroscopically derived gravities to those derived
from isochrone analysis for the sample of stars from VF05
which have effective temperatures similar to \xon.  Stars with
[M/H]$< 0.04$ are shown as blue squares and those with [M/H]$> 0.04$ are
shown with red circles.  
The green dashed line in Figure 10 is the line of equality.
The color version of the figure appears in the
electronic version of the paper.
\label{fig:loggcheck}}
\end{figure}

The above studies typically use spectroscopically
determined effective temperatures with isochrone models to determine the
stellar gravity, implicitly assuming the gravities determined from isochrone
analyses are more accurate than the spectroscopic values.  
We can compare the gravities determined from the Y$^2$ isochrones and our
spectroscopic analysis
for stars of similar spectral type to \xon.  The sample
of stars studied by VF05 all have accurate 
{\it Hipparcos} parallax measurements, allowing them to use the isochrones 
with the spectroscopically determined $T_{eff}$ to determine the stellar
gravity.  This gravity can then be compared to the gravity VF05
derive from the spectra alone, using the same techniques
used in this paper.  Figure \ref{fig:loggcheck} shows the result of
this comparison for 79 stars with $T_{eff} > 6200$ K from VF05.
Recall $T_{eff} = 6429 \pm 50$ for \xon.  There is considerable
scatter in this figure; however, in the immediate vicinity of the 
spectroscopic gravity determined for \xon, the gravities determined from
the isochrones are lower.  This is the opposite sense to what is suggested
by the transit light curve of XO-3.  As a result, there is no clear indication in
this sample of stars that the spectroscopically determined gravity for 
\xon\ is biased low.  When making this comparison, it is appropriate to
consider the accuracy of the gravities predicted by the Y$^2$ isochrones.
Hillenbrand and White (2004) report excellent agreement (better that 3\%)
between dynamically determined stellar masses from eclipsing binaries and 
masses determined from the Y$^2$ isochrones for main sequence stars more 
massive than $\sim 0.6$ \Msun; therefore, we find no reason to doubt the
Y$^2$ isochrones appropriate for \xon.  We also note that results
from transit light curve analyses are subject to various assumptions and
potential biases that may exist in the models used for the fit.  For example,
Aufdenberg et al. (2005) confirm the prediction of Allende Prieto et al.
(2002) that 1D atmosphere models, including the ATLAS models used by 
Claret (2000) to determine limb darkening coefficients, predict too much
limb darkening at optical wavelengths, depending on the treatment of 
convection and convective overshoot in the model. In fitting light curves,
such a bias will
favor models with smaller impact parameters and smaller stellar
radii.  Therefore, until better data
can be obtained for \xon\ (a precise parallax measurement and/or more
precise transit photometry), the stellar
and hence planetary radius will remain significantly uncertain.  At this point,
taking the photometric lightcurve and the spectroscopic gravity into 
account, we estimate the radius of \xonb\ to be $1.10 < R_p < 2.11$ \Rjup\
including $1\sigma$ uncertainties on both limits.  The lower limit is almost
certainly too low however, given the results of Aufdenberg et al. (2005)
and Allende Prieto et al. (2002) on limb darkening showing
that the values we have used here are likely overestimated.

\subsection{Tidal Circularization and the Eccentricity of \xonb}
\label{sec:tidal}

As mentioned in Section \ref{sec:cf}, an interesting aspect of the orbit of \xonb\ is its 
significant eccentricity, $e = \vecc \pm \eecc$, because
most planets with orbital periods
similar to \xonb\ are believed to have already been tidally circularized
(Halbwachs et al. 2005).  
Independent of exactly
how \xonb\ arrived at its current orbit, it is interesting to consider
how long it can remain in such an orbit under the influence of tidal
circularization.  
The question of tidal circularization of close
extra-solar planets has been studied by several investigators (see 
Adams \& Laughlin 2006 and references therein).  The equations governing the
tidal circularization of extra-solar planets depend on very uncertain
planetary quality factors, $Q_p$ (for example, see discussion in Gu et al.
2003) and in some cases equally uncertain stellar quality factors, $Q_*$,
if considering tides raised on the star by the planet.  As an 
illustration of the difficulty in estimating quality factors, Mathieu
(1994) point out that theoretically determined values of $Q_*$ imply a
slow rate for close binary stars to circularize which is contradicted by
observations in stellar clusters of various ages.  As a result, many
investigators try to use the observations (of binary stars and extra-solar
planets) to empirically estimate the quality factors.  For Jupiter
mass extra-solar planets, $Q_p \sim 10^5 - 10^6$ (e.g., Gu et al. 2003;
Adams \& Laughlin 2006).  

Circularization timescales depend linearly on
$Q_p$, so we assume $Q_p = 10^6$ to get predictions at the long end of what can
currently be estimated.  Using equation (18) of Gu et al. (2003) and
equation (3) of Adams \& Laughlin (2006) to estimate the circularization
time for \xonb, we find timescales of 0.29 Gyr and 0.33 Gyr, respectively,
using values corresponding to the large radius (1.9 \Rjup) for \xonb.
The circularization timescale in both studies depends on the reciprocal of
the planetary radius
to the fifth power, so the circularization timescale grows to 2.92 Gyr and 3.32
Gyr, respectively, using values corresponding to the small radius (1.25
\Rjup) for \xonb.  Clearly, there is considerable uncertainty in the value
of the circularization time; however, because circumstellar disks appear to
be lost after 10-20 Myr (e.g., Haisch et al. 2001; Mamajek et al. 2004), it 
appears that \xonb\ perhaps should have circularized by now if it originally
arrived at its current location while the circumstellar disk was still in 
place (i.e. through migration).  If instead, \xonb\ has been scattered in
to its current location (e.g., Ford \& Rasio 2007) more recently, there may
not have been enough time for circularization to occur, particularly for
the parameters corresponding to the smaller radius for \xonb.  A third
possibility is that additional, undetected planets orbiting \xon\
maintain the eccentricity of \xonb\ (e.g., Adams \& Laughlin 2006).  It is
interesting to note though that for planets in short period, eccentric
orbits such as \xonb, the tides raised in the planet can deposit
substantial energy into the planet (e.g., Gu et al. 2003; Adams \& 
Laughlin 2006) which can inflate it to radii coresponding to
the larger value we measure (Gu et al. 2003).  It therefore appears
\xonb\ can again serve as a very interesting benchmark for studies of tidal
circularization and tidal heating once a more accurate radius can be
established for the planet.

\section{Summary}

\xonb\ is a massive planet or low mass brown dwarf in a short period ($3.2$
d), eccentric ($e=0.26$) orbit, around a somewhat evolved F5 star.  There is 
relatively large uncertainty in the planetary parameters owing to the unknown 
distance to \xon.  The mass of \xonb\ is 11.57 -- 13.97 \Mjup\ and the radius
is 1.10 -- 2.11 \Rjup.  The larger mass and radius are favored by our
spectroscopic analysis of \xon, and the smaller values are favored by
the light curve analysis.  A precise trigonometric parallax measurement
or more accurate photometric light curve date are needed to distinguish 
between these values.

\acknowledgments

We wish to thank the anonymous referee for many useful comments which
improved the manuscript.
The University of Hawaii staff have made the operation on Maui possible; we
thank especially Bill Giebink, Les Hieda, 
Jake Kamibayashi,
Jeff Kuhn, Haosheng Lin, Mike Maberry,
Daniel O'Gara, 
Joey Perreira, Kaila Rhoden, and the director of the IFA, Rolf-Peter Kudritzki.

This research has made use of a Beowulf cluster constructed by Frank
Summers; the SIMBAD database, operated at CDS, Strasbourg, France;
data products from the Two Micron All Sky Survey (2MASS) and
the Digitized Sky Survey (DSS);
source code for transit light-curves (Mandel \& Agol 2002);
and community access to the 
Hobby-Eberly Telescope (HET), which is a joint project of the University
of Texas at Austin, the Pennsylvania State University, Stanford
University, Ludwig-Maximilians-Universit\"{a}t M\"{u}nchen, and
Georg-August-Universit\"{a}t G\"{o}ttingen.  The HET is named in honor
of its principal benefactors, William P. Hobby and Robert E. Eberly.
We thank the HET night-time and day-time support staff.
We also wish to thank the support staff at McDonald Observatory, and to
especially thank M. Huerta for his assistance observing at the 2.7 m HJS
Telescope.

XO is funded primarily by NASA Origins grant 
NNG06GG92G and the Director's Discretionary Fund of STScI.
C.~M. Johns--Krull and L. Prato wish to ackowledge partial support from 
NASA Origins of Solar Systems grant 05-SSO05-86.
We thank Brian Skiff and Josh Winn for noting that a preprint had an error in the coordinates and there was a sign error in our radial velocity fitting algorithm, respectively; both were corrected prior to publication.

\clearpage

\begin{deluxetable}{cccccc}
\tabletypesize{\small}
\tablewidth{0pt}
\tablecaption{{\rm XO Survey \& E.T.\ Light Curve Data}\tablenotemark{a}}
\startdata
\hline
\hline
Heliocentric Julian Date & Light Curve & Uncertainty & Filter & N\tablenotemark{b} & Observatory \\
                         &  [mag]      & (1-$\sigma$) [mag] & & & \\
\hline
2452932.10913 & -0.0017 & 0.0027  &   W\tablenotemark{c} &  1 &  XO \\
2452932.10938 & -0.0032 & 0.0026  &   W\tablenotemark{c} &  1 &  XO \\
2452932.11597 &  0.0024 & 0.0026  &   W\tablenotemark{c} &  1 &  XO \\
2452932.11621 &  0.0021 & 0.0025  &   W\tablenotemark{c} &  1 &  XO \\
2452932.12305 & -0.0016 & 0.0026  &   W\tablenotemark{c} &  1 &  XO \\
\hline
\enddata
\tablenotetext{a}{The complete version of this table is in the electronic edition of
the Journal.  The printed edition contains only a sample.}
\tablenotetext{b}{Average of N measurements}
\tablenotetext{c}{The filters used in the XO project telescopes are
described in McCullough et al. (2005).  The bandpass is essentially
flat from 4000 -- 7000 \AA.}
\label{table:lc}
\end{deluxetable}

\begin{deluxetable}{ccccc}
\tabletypesize{\small}
\tablewidth{0pt}
\tablecaption{{\rm All-sky Photometric Magnitudes}}
\startdata
\hline
\hline
Star\tablenotemark{a} & B     & V     & ${\rm R_C}$ & ${\rm I_C}$ 	\\
\hline
\xon\ 	       & 10.25 &  9.80 &  9.54       &  9.28 \\
   1           & 11.37 & 10.82 & 10.50       & 10.18 \\
   2           & 12.38 & 11.62 & 11.17       & 10.76 \\
   3           & 14.81 & 13.07 & 12.17       & 11.30 \\
   4           & 10.13 &  9.03 &  8.43       &  7.88 \\
   5           & 12.67 & 11.87 & 11.43       & 10.92 \\
   6           &  9.26 &  8.78 &  8.50       &  8.21 \\
   7           & 15.52 & 13.46 & 12.41       & 11.50 \\
   8           & 15.34 & 13.84 & 12.59       & 11.51 \\
\enddata
\tablenotetext{a}{Stars are identified in Figure \ref{fig:allsky}.}
\label{table:allsky}
\end{deluxetable}

\begin{deluxetable}{lcl}
\tabletypesize{\small}
\tablewidth{0pt}
\tablecaption{{\rm The Star \xon}}
\startdata
\hline
\hline
Parameter & Value & Reference\\
\hline
RA (J2000.0) & $ 04^h21^m52^s.71 $ & a,b \\
Dec (J2000.0) & +57\arcdeg49\arcmin01\arcsec.9 & a,b \\
$V, V_T, V_{calc}$ & 9.80$\pm$0.03,$9.904\pm0.027$,$9.86\pm0.027$ & c,b,d\\
$B-V, B_T-V_T, (B-V)_{calc}$ & 0.45$\pm$0.04,$0.451\pm0.040$,$0.383\pm0.034$ & c,b,d\\
$V-R_C$ & 0.26$\pm$0.04 & c\\
$R_C-I_C$ & 0.26$\pm$0.06,& c\\
$J$   & $9.013\pm0.029$ & e\\
$J-H$ & $0.168\pm0.034$ & e\\
$H-K_{\rm s}$ & $0.054\pm0.026$ & e\\
Spectral Type & \sptype & c \\
d & $\vDs \pm \eDs$ pc & c \\
$(\mu_\alpha,\mu_\delta)$ & ${\rm (-0.6 \pm 2.7, 1.6 \pm 2.6)~mas~yr^{-1}}$ & b \\
GSC 	& 03727-01064 & a \\
\enddata
\tablerefs{\\
a) SIMBAD\\
b) Tycho-2 Catalogue, H{\o}g et al (2000) \\
c) this work \\
d) Calculated from Tycho-2 measurements using standard transformations\\
e) 2MASS, Skrutskie e al. (2006)
}
\label{table:star}
\end{deluxetable}
                                                                                            
\begin{deluxetable}{cccccc}
\tabletypesize{\small}
\tablewidth{0pt}
\tablecaption{{\rm Radial Velocity Shifts}}
\startdata
\hline
\hline
Heliocentric& Radial Velocity &  Uncertainty & $O-C$ & Bisector & \\
Julian Date &  Shift [\mps] &  (1 $\sigma$) [\mps]  & [\mps] & Span [\mps] & Telescope \\
\hline
    2454037.0112 &     1265 &    209 &  107 & -31 & HJS \\
    2454038.0098 &     2001 &    216 &  174 &   4 & HJS \\
    2454137.8215 &        0 &    134 &  -92 &   0 & HJS \\
    2454138.8124 &      216 &    134 & -112 & 170 & HJS \\
    2454139.8128 &     2625 &    139 & -117 & 133 & HJS \\
    2454140.8131 &      267 &    139 &  -47 & 203 & HJS \\
    2454141.8047 &       77 &    149 &   13 & 112 & HJS \\
    2454142.7967 &     2761 &    139 &   -1 &  68 & HJS \\
    2454143.7999 &      881 &    143 &  242 &  42 & HJS \\
    2454144.7967 &      -35 &    161 &   34 & 152 & HJS \\
\\
    2454005.8587 &     1454 &    129 &  103 & \nodata & HET \\
    2454006.8597 &     -976 &    122 &   16 & \nodata & HET \\
    2454113.5842 &      -71 &    184 &   44 & \nodata & HET \\
    2454121.7100 &     -981 &    188 &  -43 & \nodata & HET \\
    2454121.7190 &     -965 &    204 &  -16 & \nodata & HET \\
    2454122.5635 &    -1152 &    182 &   52 & \nodata & HET \\
    2454127.7046 &     -240 &    195 &   32 & \nodata & HET \\
    2454128.7048 &    -1271 &    183 &   21 & \nodata & HET \\
    2454158.6220 &     1191 &    144 &  -32 & \nodata & HET \\
    2454159.6091 &     -278 &    132 &  -31 & \nodata & HET \\
    2454162.6193 &      -30 &    194 & -248 & \nodata & HET \\
\enddata
\label{table:rvobs}
\end{deluxetable}

\begin{deluxetable}{lccc}
\tabletypesize{\small}
\tablewidth{0pt}
\tablecaption{{\rm Transit Timing Measurements}}
\startdata
\hline
\hline
  &  &  & Transit Midpoint  \\
 UT Date & Observer & Filter & (HJD - 2450000.0)  \\
\hline
2003 Dec 24 & XO & W\tablenotemark{a} & 2997.72729 \\
2004 Oct 22 & XO & W\tablenotemark{a} & 3300.90479 \\
2004 Nov 10 & XO & W\tablenotemark{a} & 3320.05713 \\
2004 Dec 12 & XO & W\tablenotemark{a} & 3351.99341 \\
2006 Sep 21 & CF & R & 3999.86011 \\
2006 Sep 21 & MF & R & 3999.85913 \\
2006 Oct 16 & TV & R & 4025.39673 \\
2006 Oct 20 & EM & V & 4028.58984 \\
2006 Oct 23 & CF & B & 4031.77759 \\
2006 Oct 23 & CF & R & 4031.77954 \\
2006 Oct 23 & CF & V & 4031.78052 \\
2006 Nov 08 & CF & B & 4047.73022 \\
2006 Nov 08 & CF & R & 4047.73804 \\
2006 Nov 08 & CF & V & 4047.73340 \\
2006 Nov 24 & EM & I & 4063.69019 \\
2006 Nov 24 & MF & I & 4063.69897 \\
2006 Nov 27 & MF & I & 4066.88525 \\
2007 Jan 26 & EM & I & 4127.53076 \\
2007 Feb 15 & CF & V & 4146.67725 \\
2007 Mar 19 & MF & R & 4178.58936 \\
\enddata
\tablenotetext{a}{The filters used in the XO project telescopes are
described in McCullough et al. (2005).  The bandpass is essentially
flat from 4000 -- 7000 \AA.}
\label{table:ttime}
\end{deluxetable}

\begin{deluxetable}{lcl}
\tabletypesize{\small}
\tablewidth{0pt}
\tablecaption{{\rm Orbital Solution and Planetary Parameters \xonb}}
\startdata
\hline
\hline
Parameter & Value & Notes\\
\hline
$P $ 		& \vperiod$\pm$\eperiod\  d                     & \\
$t_c $	 	& \vjd$\pm$\ejd\          HJD                   & \\
$e$             & \vecc$\pm$\eecc                               & \\
$\omega$        & $-15.4\pm6.6$          degrees               & \\
$T_\circ$       & 2454024.7278$\pm$0.0570  HJD                  & \\
$K $ 		& \vrvK$\pm\ervK$\        \mps                  & \\
$M_{\rm p}{\rm sin}i$   & \vMp$\pm$\eMp                         & b\\
$R_{\rm p}/R_{\rm s} $ 	& (0.92$\pm$0.04) $\times$ \Rjup/\Rsun 	& a\\
$i $ 			& \vincl$\pm$\eincl\ deg		& b,c\\
$a $ 			& \vap$\pm$\eap\ A.U. 			& b \\
$M_{\rm p} $ 		& \vMptot$\pm$\eMptot\ \Mjup		 	& b,d\\
$R_{\rm p} $ 		& \vRp$\pm$\eRp\ \Rjup			& a,b,c\\
\enddata
\tablecomments{\\
a) \Rjup\ = 71492 km, i.e. the equatorial radius of Jupiter\\
b) for $M_*$ = \vMs$\pm$\eMs\ \Msun \\
c) for $R_*$ = \vRs$\pm$\eRs\ \Rsun \\
d) \Mjup\ = 1.8988e27 kg 
}
\label{table:planet}
\end{deluxetable}

\begin{deluxetable}{ccc}
\tabletypesize{\small}
\tablewidth{0pt}
\tablecaption{{\rm Results of the SME Analysis}}
\startdata
\hline
\hline
          &      &  Precision \\
Parameter & Mean &  (1 $\sigma$)	\\
\hline
$T_{eff} [K]$	&6429 	&   50		\\
log$g$ [\cmpss]	&3.95  	&   0.062	\\
V~sin~$i$ [\kps] &18.54 &   0.17	\\
\ [M/H]     	&-0.204	&   0.023	\\
\ [Na/H]	&-0.346	&   0.046	\\
\ [Si/H]	&-0.171	&   0.017	\\
\ [Ti/H]	&-0.262	&   0.049	\\
\ [Fe/H]	&-0.177 &   0.027	\\
\ [Ni/H]	&-0.220	&   0.033	\\
\ [Si/Fe]   	&0.006	&   0.032	\\
\enddata
\label{table:sme}
\end{deluxetable}

\begin{deluxetable}{lccc}
\tabletypesize{\small}
\tablewidth{0pt}
\tablecaption{{\rm Spectroscopically Derived Stellar parameters}}
\startdata
\hline
\hline
Parameter	&	@ 240 pc&	@ 260 pc&	@ 280 pc\\
\hline
   		&	1.33	&	1.36	&	1.39	\\
Mass [\Msun]   	&	1.36	&	1.41	&	1.43	\\
    		&	1.39	&	1.44	&	1.48	\\
    		&	    	&	    	&	    	\\
    		&	1.92	&	2.08	&	2.22	\\
Radius [\Rsun] 	&	1.95	&	2.13	&	2.27	\\
    		&	1.99	&	2.17	&	2.34	\\
    		&	    	&	    	&	    	\\
    		&	3.98	&	3.93	&	3.87	\\
Log(g) [\cmpss] &	4.00	&	3.95	&	3.89	\\
    		&	4.02	&	3.97	&	3.91	\\
    		&	    	&	    	&	    	\\
    		&	2.66	&	2.53	&	2.41	\\
Age [Gyr]     	&	2.78	&	2.69	&	2.68	\\
    		&	3.06	&	2.83	&	2.96	\\
\enddata
\tablecomments{
For each parameter, the middle row is the maximum likelihood value, and the
values in the rows above and below span the 68\% likelihood of the probability
distributions (cf. Figure \protect{\ref{fig:stellarparam}}). The three columns correspond to three assumed
distances for \xon.}
\label{table:stellarparam}
\end{deluxetable}

\clearpage

\clearpage
\thispagestyle{empty}
\setlength{\voffset}{-10mm}
\clearpage
\setlength{\voffset}{0mm}

\end{document}